\documentclass[acmconf,authorversion,manuscript]{acmart}

\AtBeginDocument{%
  \providecommand\BibTeX{{%
    Bib\TeX}}}

\setcopyright{acmlicensed}
\copyrightyear{2018}
\acmYear{2018}
\acmDOI{XXXXXXX.XXXXXXX}

\acmConference[Conference acronym 'XX]{Make sure to enter the correct
  conference title from your rights confirmation emai}{June 03--05,
  2018}{Woodstock, NY}
\acmISBN{978-1-4503-XXXX-X/18/06}

\usepackage{subcaption}
\usepackage{graphicx}

\begin{document}


\title[Another Body in the World]{Another Body in the World: Flusserian Freedom in Mixed Reality}

\author{Aven-Le ZHOU}
\email{aven.le.zhou@gmail.com}
\orcid{0000-0002-8726-6797}
\affiliation{%
  \institution{The Hong Kong University of Science and Technology (Guangzhou)}
  \streetaddress{No.1 Du Xue Rd, Nansha District}
  \city{Guangzhou}
  \state{Guangdong}
  \country{P.R.China}
}

\author{Lei XI}
\email{xiju524954135@gmail.com}
\orcid{0009-0000-1802-6901}
\affiliation{%
  \institution{University of Art and Design Linz}
  \city{Linz}
  \country{Austria}
}

\author{Kang Zhang}
\orcid{0000-0003-3802-7535 }
\affiliation{%
  \institution{The Hong Kong University of Science and Technology (Guangzhou)}
  \streetaddress{No.1 Du Xue Rd, Nansha District}
  \city{Guangzhou}
  \state{Guangdong}
  \country{P.R.China}
}

\begin{teaserfigure}
  \centering
  \includegraphics[width=\textwidth]{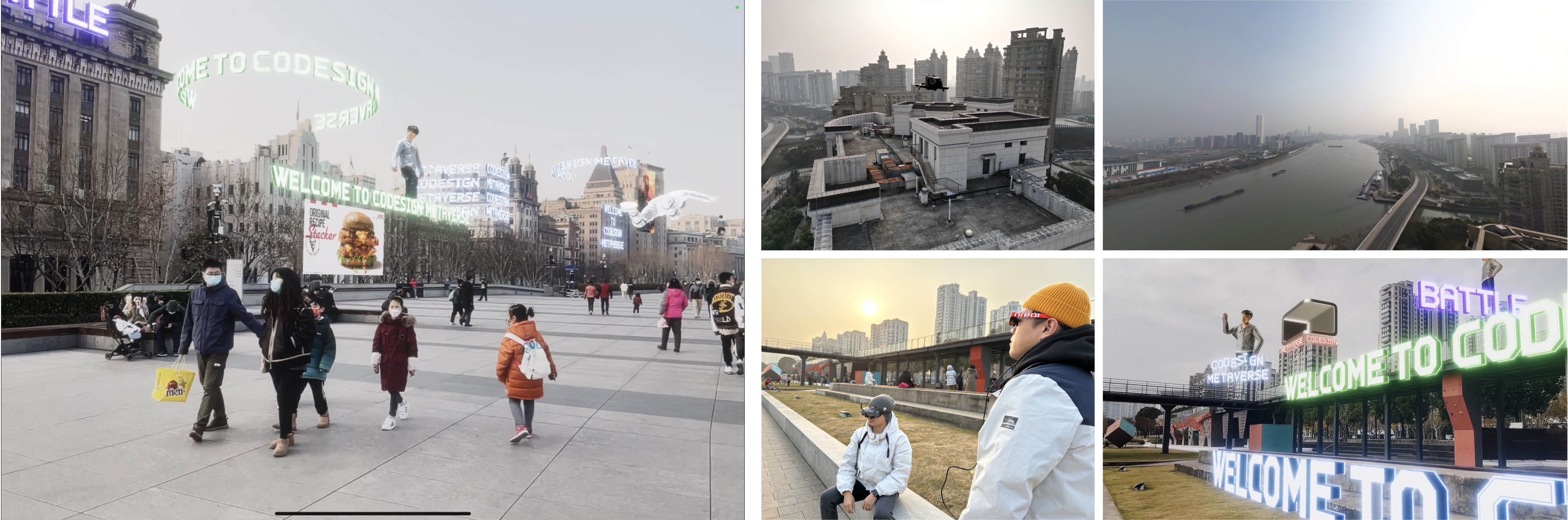}
  \caption{Demonstration of Surrealism Me across multiple sites.}
\end{teaserfigure}

\renewcommand{\shortauthors}{Zhou and Xi et al.}

\begin{abstract}
In Flusserian view of media history, humans often misperceive the world projected by media to be the world itself, leading to a loss of freedom. This paper examines Flusserian Freedom in the context of Mixed Reality (MR) and explores how humans can recognize the obscuration of the world within the media (i.e., MR) and understand their relationship. The authors investigate the concept of playing against apparatus and deliberately alienating the perception of the projected world through an artwork titled “Surrealism Me.” This artwork enables the user to have another body within MR through interactive and immersive experiences based on the definition of Sense of Embodiment. The purpose of this work is to raise awareness of the domination of media and to approach Flusserian freedom within contemporary technical arrangements.
\end{abstract}

\ccsdesc[500]{Applied computing~Media arts}
\ccsdesc[500]{Computing methodologies~Computer graphics}


\keywords{Flusser, Media Studies, Virtual Body, Sense of Embodiment, Bodily Movement, Embodied Experience}


\maketitle

\section{Introduction}

Czech-Brazilian philosopher Vilém Flusser claims media are things that mediate between humans and the world \cite[p.~9]{flusser2000}. The world is not "immediately accessible" to humans, so they need to "make it comprehensible" through media. Humans invent media to help them make sense of the chaotic world \cite{ieven2003} and orient themselves in it \cite[p.~10]{flusser2000}. However, media do not always necessarily offer faithful representations of the world but rather encode the information they present in particular ways.  Since humans often misperceive the world projected by media to be the world itself, media becomes "screens" or "coverings" of the world \cite[p.~2]{flusser2013}, namely, obstacles between humans and the world. This misperception hinders humans from orienting themselves in the world through media and leads to a loss of freedom.

In the Flusserian view of media history, such loss of freedom occurs frequently. With the development of traditional images, such as painting and drawing, humans forget that these images are ways of orienting themselves in the world and believe them to be reliable representations of the world \cite[p.~10]{flusser2000}. To overcome the obscuration of the world by traditional images, humans invent written texts \cite[p.~74]{flusser1997}. However, since written texts are "one step further away from concrete experience than images," they often lead to "a symptom of a bigger alienation" \cite[p.~25]{flusser2002}. When the domination of written texts peaks in the nineteenth century, the technical images emerge, exemplified by the invention of photography \cite{ieven2003}. The apparatuses that make the technical images are often black boxes to the ordinary user. In other words, users often do not understand how the world is encoded by apparatuses. Apparatuses cease to be instruments that make the world accessible and humans "finally become a function of the images they create" \cite[p.~10]{flusser2000}. 

The call for freedom is significant for Flusser's media theory  \cite{ieven2003}. Flusser suggests that humans can still be free by mastering and reinstructing the media that dominate us \citetext{\citealp[p.~73]{flusser2011}; \citealp[pp.~81-82]{flusser2000}}. According to Flusser, freedom means playing against the apparatus \cite[p.~80]{flusser2000}. More specifically, it involves imposing human intentions into the program of apparatus, exhausting its potential \cite{poltronieri2014communicology}. In this process, the user consciously creates information that is unpredictable to the apparatus, adding things that are not in the program \cite[pp.~81-82]{flusser2000}. In doing so, the user is not merely a function of the apparatus but conditioned on apparatus and vice versa, which opens the way to freedom. 



Over the decades, technical images have evolved from photography to more immersive digital imaging technologies. Although writings \cite[pp.~39,148]{popiel2012vilem} based on Flusserian thoughts address the relationship between new digital imaging technologies and freedom, they often only involve Virtual Reality (VR) while not mentioning Mixed Reality (MR). The tech industry defines MR as a technology that "brings together the real world and digital elements," which indicates that the reality in MR is often misrecognized as the real world. But it's actually a representation of the world projected through the digital apparatus. Thus, a problem arises: just as a medium is often misperceived to be the real world, humans tend to regard the technical images of MR as the real world and are thus in danger of becoming a function of apparatus.

Our work mainly investigates MR, the current misperceived medium, leading the user to recognize the obscuration of the world in the context of MR and to better understand the relationship between media (i.e., MR) and the world. Thus, we explore the concept of playing against apparatus and deliberately alienating the perception of the projected world in MR. Through various embodied experiences and interactions in an artwork entitled Surrealism Me, we aim to open up space for freeing humans from the domination of digital media. Surrealism Me enables the user to have a virtual body in MR through (1) tasking the artificial intelligence (AI) algorithms to mimic and adapt to the user's unpredictable bodily (and facial) movements, in which the user plays against the apparatus, and (2) intentionally further manipulating the user's visionary sensation to reveal that reality in MR is not the real world. The embodied experiences and multi-model interactions inspire a concern for freedom in the contemporary context of digital technical arrangements.

\section{Method}

\subsection{Playing against Apparatus}\label{play}

Flusser proposes imposing unpredictable human intentions into the apparatus and exhausting its program as an important way to play against it (thus human can be free) \cite [pp.~80-82] {flusser2000}. For instance, game playing is the typical practice of humans playing against apparatus \cite{poltronieri2014communicology}, in which the player tends to exhaust the program within the game by constantly undertaking unpredictable new actions \cite[pp.26-27]{flusser2000}. The end of unpredictability means going beyond the limitation of the program and forcing the apparatus to enhance its potential. The interactions between the user and the AI share similarities with the ones between the player and the game, in which humans and apparatus mutually condition each other. The user constantly provides new data (that's beyond AI algorithm's limitation) to train the neural network (for motion generation) to evolve, thus plays against the algorithm (as program of the AI). Our work allows the user to apply their unpredictable intentions (i.e., new motion data) to play against the AI's program which address the foundation to discuss about the Flusserian freedom.

\subsection{Having Another Body in Mixed Reality}\label{anotherbody}

The dissociation of our body from ourselves is not impossible, and the "application of the right sensory input at the right time" can lead to the illusion of owning another body \cite{9495125}. Kilteni and Slater define the feeling of having a body as the Sense of Embodiment (SoE)\label{SOE}, which encompasses three sensations: (1) being the author of the body's movements, (2) experiencing sensations originating from the body, and (3) being located inside the body \cite{6797786}. Research on SOE in Virtual Reality (VR) and Augmented Reality (AR) investigates the psychological effects of appearance, interaction, and sensory feedback in creating a sense of owning a virtual body \cite{9495125, 7504682}. VR and AR allow one to experience a different embodied form that can vary in shape, size, or appearance \cite{9495125}. Continuing this context, our work enables the user to place their virtual body in the MR projected world with embodied experiences to address the discussion on Flusserian Freedom in media studies.

\begin{figure}[b]
  \centering
  \begin{subfigure}{0.495\textwidth}
      \includegraphics[width=\textwidth]{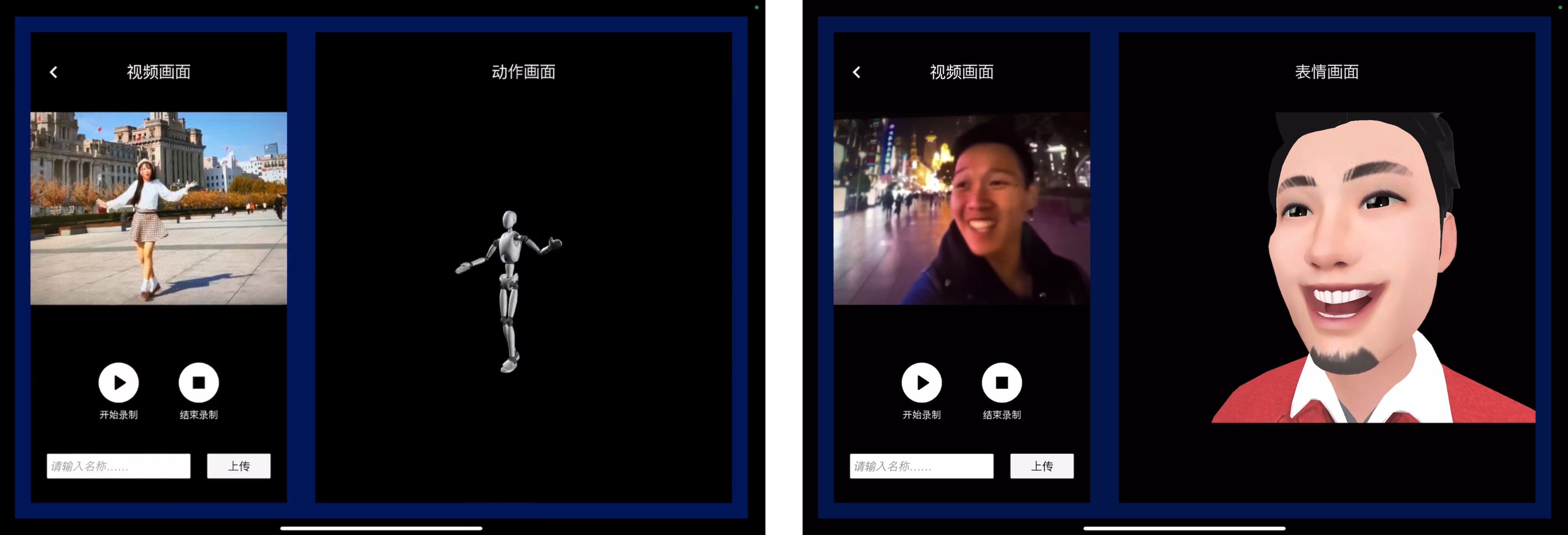}
      \caption{Interface to capture the user's bodily \& facial movement.}
      \label{fig:imitate-a}
  \end{subfigure}
  \hfill
\begin{subfigure}{0.495\textwidth}
  \includegraphics[width=\textwidth]{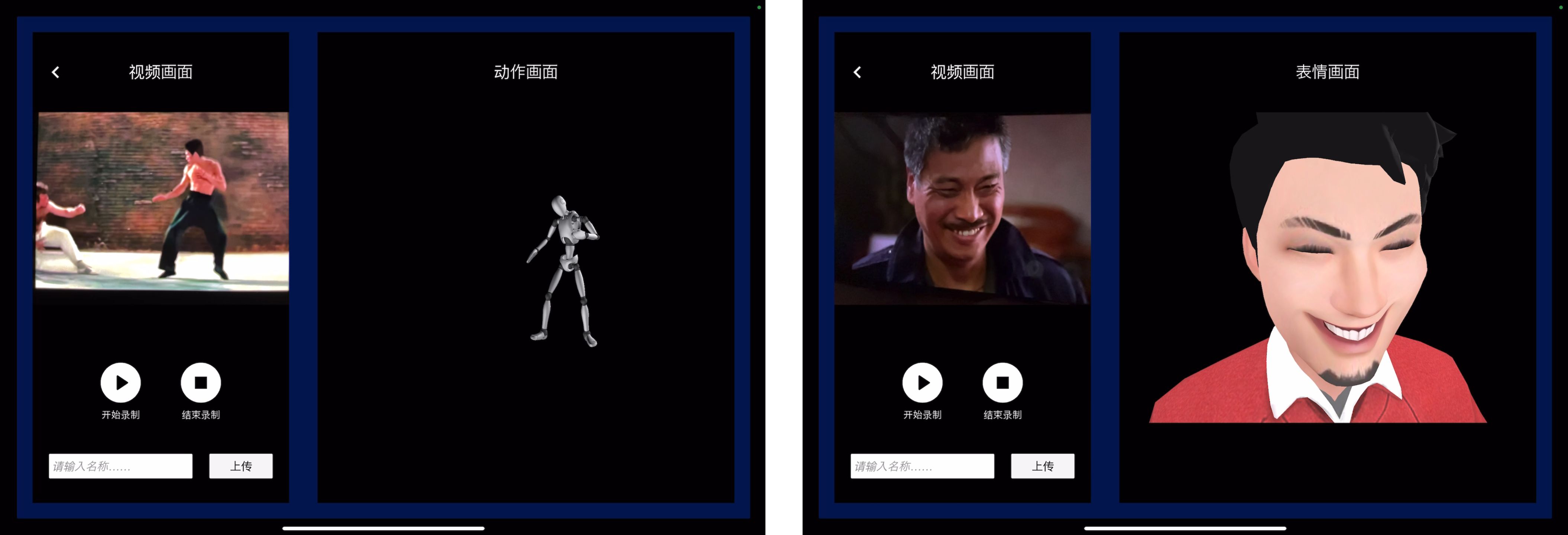}
  \caption{Interface to create movement from user-contributed video.}
  \label{fig:imitate-b}
\end{subfigure}
  \caption{Imitating the user's body.}
  \label{fig:imitate}
  \end{figure}

\section{Experiment: Surrealism Me}

To critically explore the misperception of the technical images in contemporary digital contexts, we primarily investigate MR and how humans access and understand the media and the world. We focus on establishing a virtual body in MR with enhanced embodied experiences following the Sense of Embodiment (SOE) definitions. Our artwork, Surrealism Me, facilitates embodied interaction and perception for the user to place their virtual body in MR, ultimately allowing them to attain a sense of freedom within these technical images. This endeavor allows the user to manipulate the bodily movements of the virtual body, perceive visionary sensations, and immerse oneself within the virtual body. 

\subsection{Embodied Experience through Multi-Model Interaction} \label{manipulation}

To address Flusser's concern about humans becoming a function of the technical images \cite[pp.10]{flusser2000} by "imposing unpredictable human intentions into the programs of apparatuses" \cite{poltronieri2014communicology}, we explore different models of interaction to enhance the user's embodied experience of the virtual body in MR. In Surrealism Me, we task the computer to imitate humans from different sources (i.e., the user's bodily movement, user-contributed content, and open-source motion dataset) and let the virtual body be a puppet (a user self-created avatar, see explaination in Appendix). We utilize multiple AI algorithms to capture and generate motion data to drive the virtual body. Surrealism Me stores the captured motion into an integrated database along with the open-source data. On the other hand, we constantly apply AI to generate new motion with the database as training set. The database keeps expanding with the engagements of humans and AI. 


\subsubsection{Imitating User's Body}

Fig.\ref{fig:imitate-a} illustrates the interface of Surrealism Me and indicates the user's bodily movements and facial expressions as the source for capturing, establishing a direct and fully-connected control between the user and the virtual body. We employ pose estimation and facial expression recognition algorithms by using the Unity AR Foundation package \cite{unityAR} to capture and record the user's bodily and facial movements as motion data. Additionally, we go beyond mere mimicry; with user-contributed content, we apply the same techniques to extract motion and emotion data from the provided footage. For instance, a user can select a favorite Bruce Lee movie clip and apply the Kung Fu movements to their virtual body, rather than solely relying on their own bodily movement (see Fig.\ref{fig:imitate-b}).



\begin{figure}[b]
  \centering
  \begin{subfigure}{0.3\textwidth}
      \includegraphics[width=\textwidth]{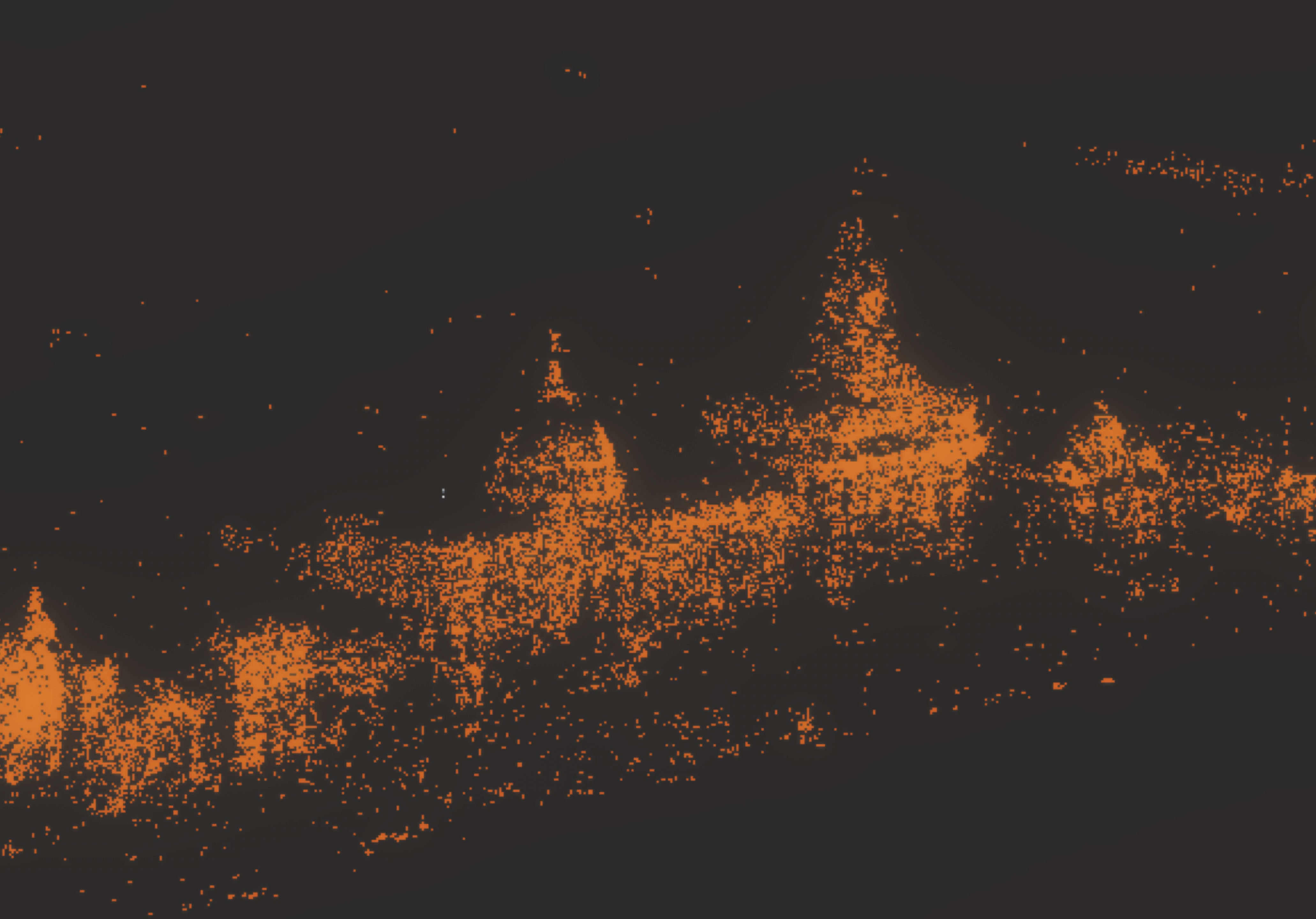}
      \caption{3D point-clouds of selected Area of Interest (AOI).}
      \label{fig:MRtech-a}
  \end{subfigure}
  \hfill
  \begin{subfigure}{0.3\textwidth}
      \includegraphics[width=\textwidth]{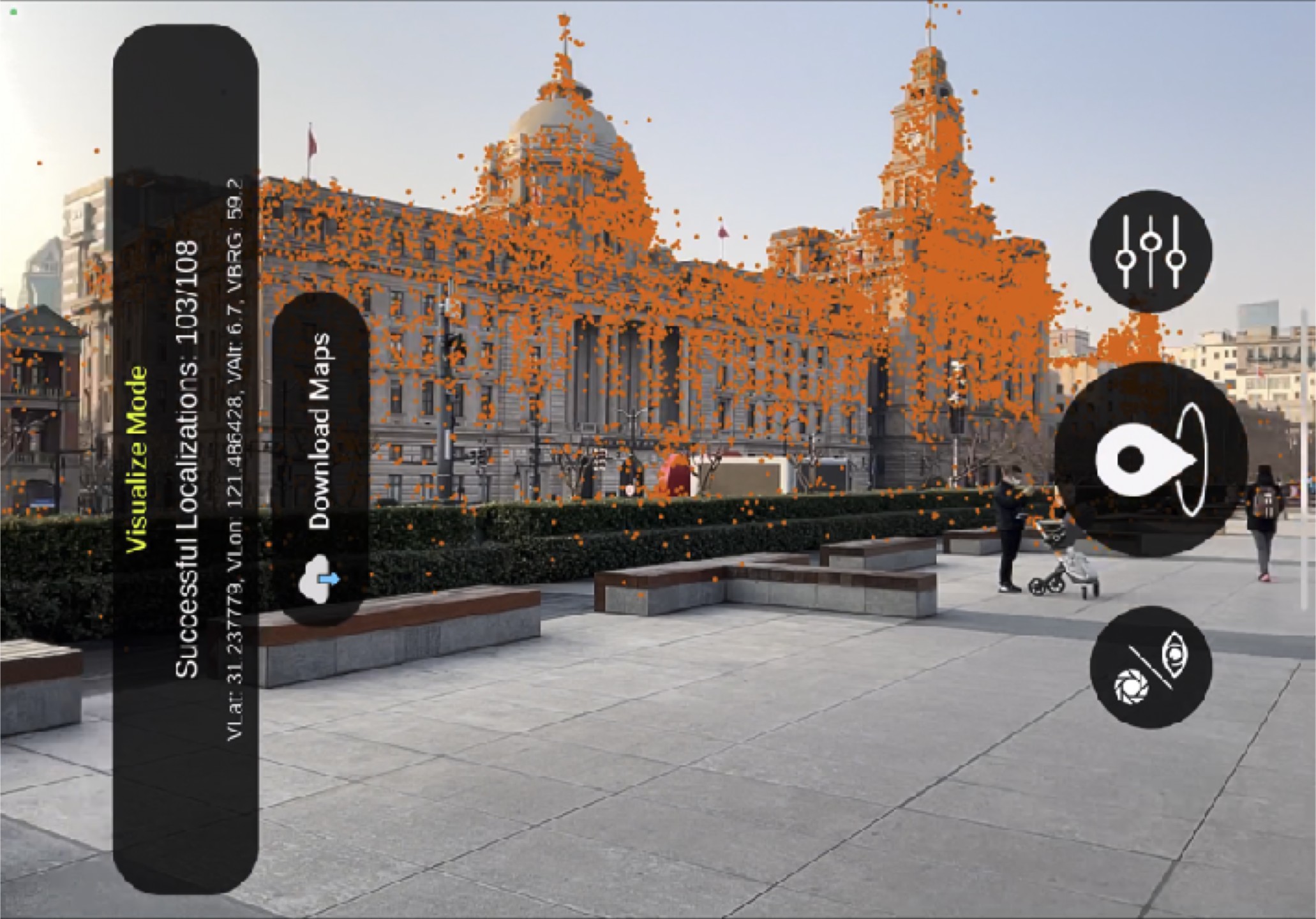}
      \caption{AOI identification with VPS for localization.}
      \label{fig:MRtech-b}
  \end{subfigure}
  \hfill
  \begin{subfigure}{0.3\textwidth}
      \includegraphics[width=\textwidth]{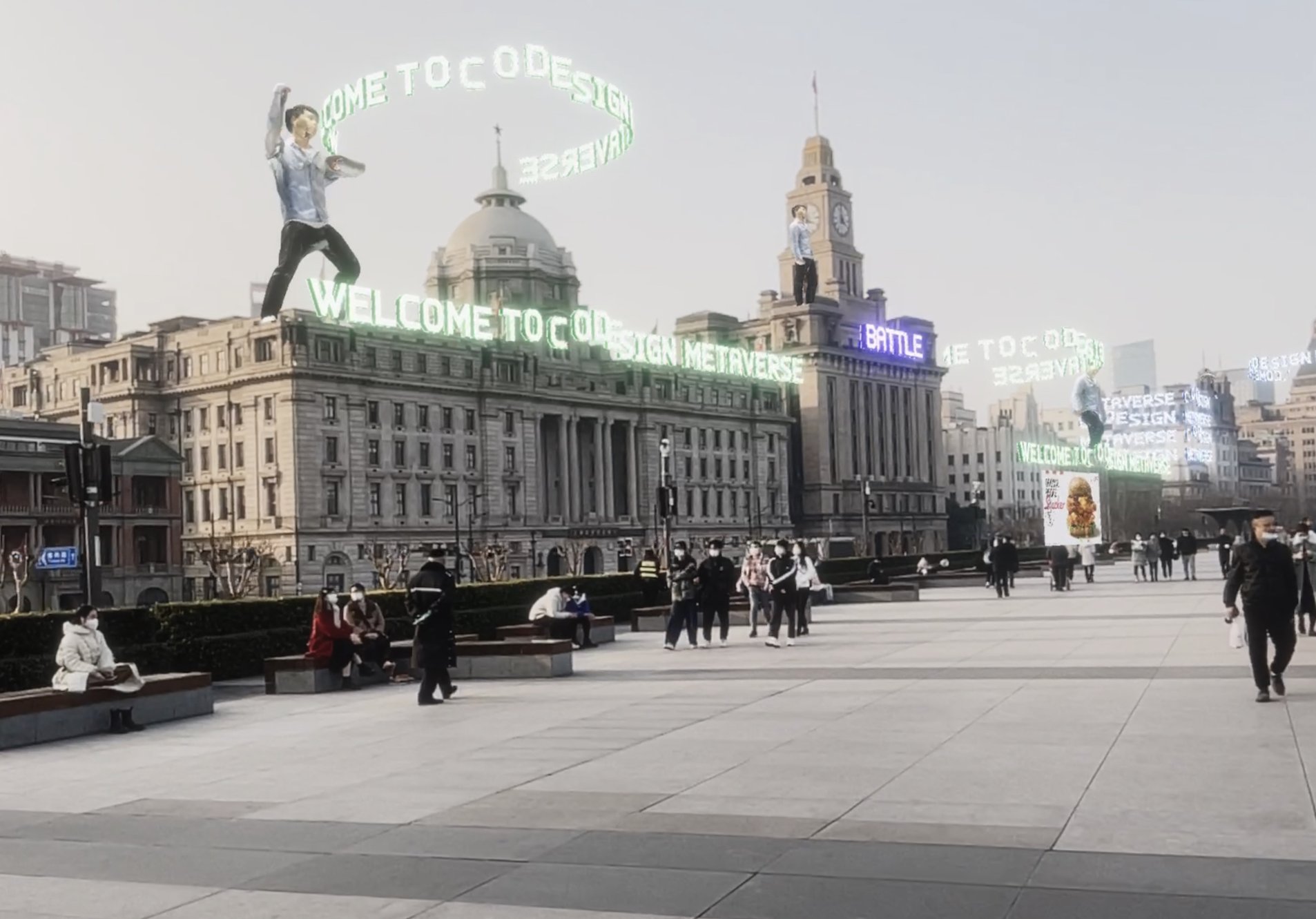}
      \caption{Virtual content positioned in the MR environment.}
      \label{fig:MRtech-c}
  \end{subfigure}
  \caption{MR technology work in practice.}
  \label{fig:MRtech}
  \end{figure}

\begin{figure}[b]
  \centering
  \begin{subfigure}{0.3\textwidth}
      \includegraphics[width=\textwidth]{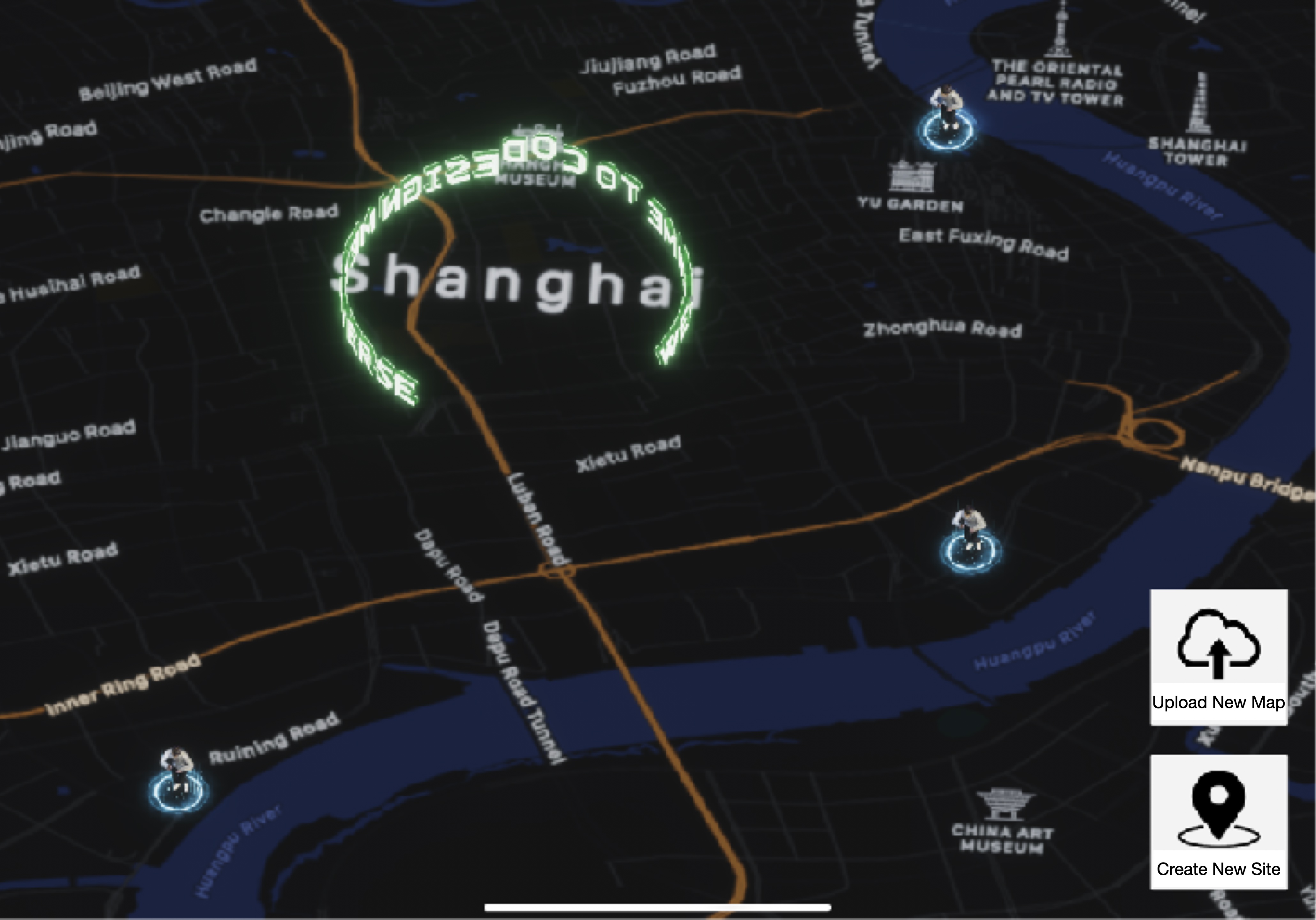}
      \caption{User-registered AOIs map and "Create New Site" button.}
      \label{fig:map-a}
  \end{subfigure}
  \hfill
  \begin{subfigure}{0.3\textwidth}
      \includegraphics[width=\textwidth]{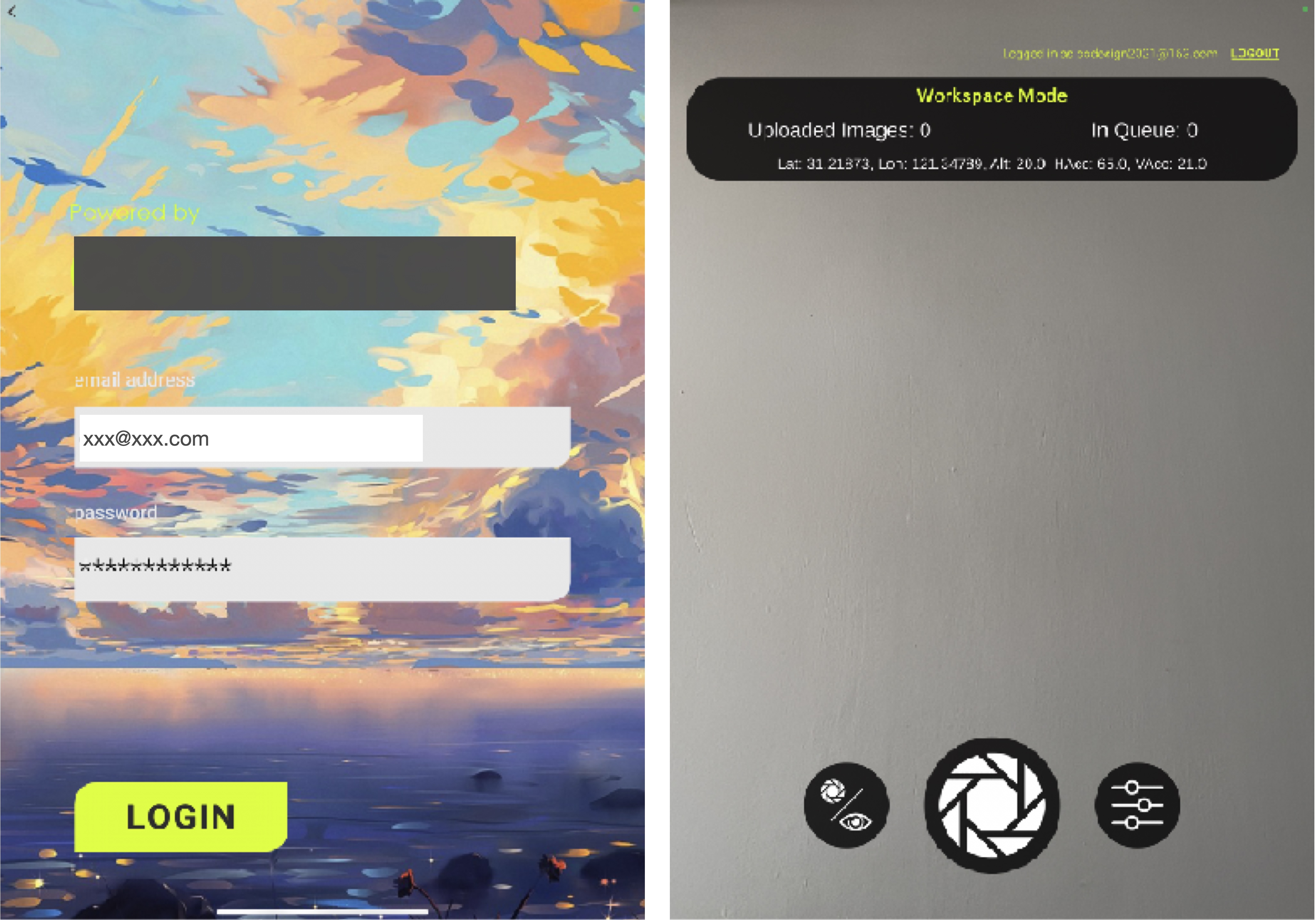}
      \caption{Interface to capture and upload photos.}
      \label{fig:map-b}
  \end{subfigure}
  \hfill
  \begin{subfigure}{0.31\textwidth}
      \includegraphics[width=\textwidth]{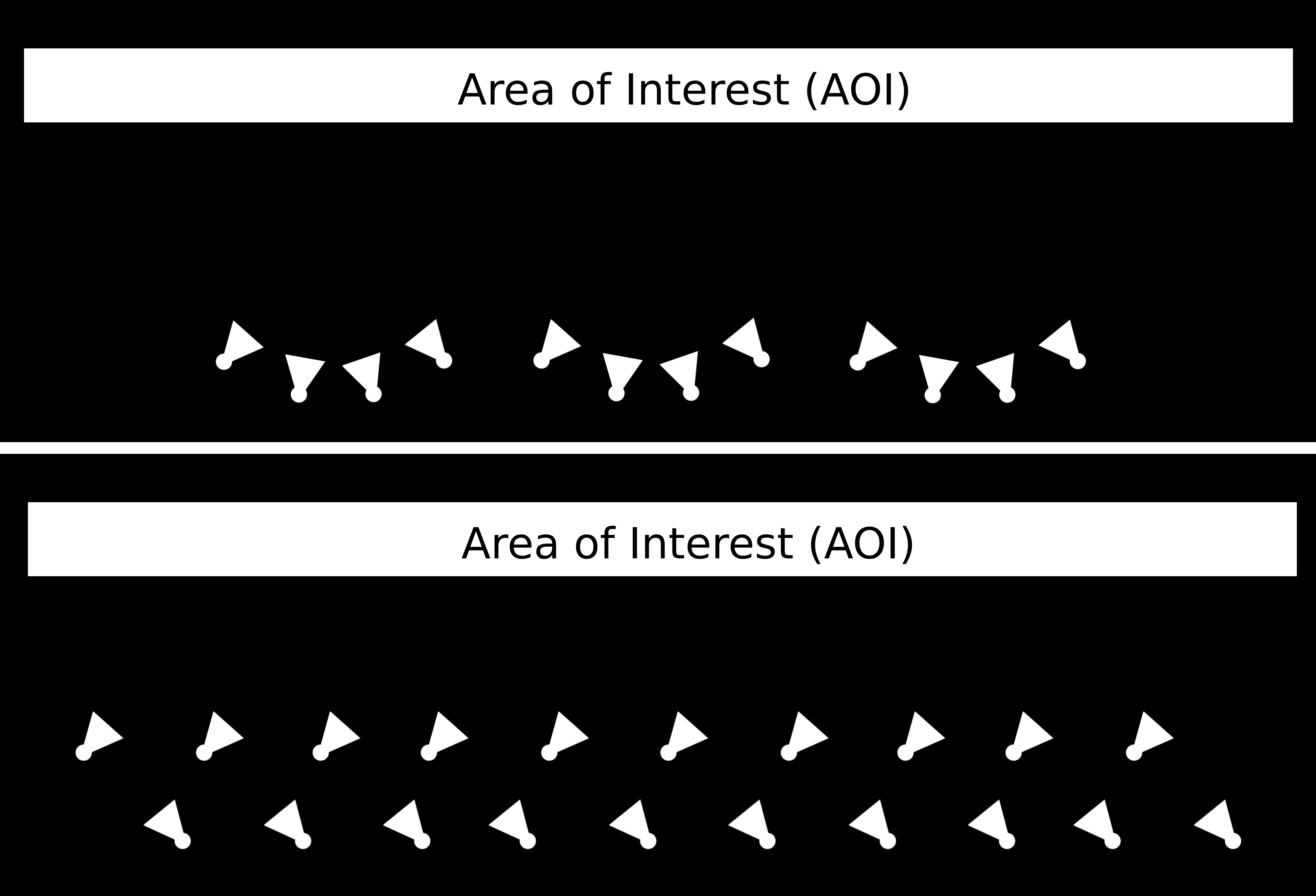}
      \caption{Two effective methods for taking photos.}
      \label{fig:map-c}
  \end{subfigure}
  \caption{Create new site function and crowd-source the infrastructure.}
  \label{fig:map}
  \end{figure}

\subsubsection{Beyond the Physical Capabilities of the User's Body}    

Moreover, we tend to establish a motion database not merely relies on the user's input. We harness the generative capability of AI to introduce semi-autonomous agency in creating movements for the virtual body. We initially incorporate an open-source dataset \cite{mixamo} into the motion database to provide clips such as dancing and extreme sports. Furthermore, we adapt and train an open-source neural network \cite{9042236} with the database (as training data) to generate new motions and expand the dataset. Additionally, we explore cross-model translation algorithms to generate motions from different modalities, such as music \cite{9042236} and text \cite{9880214}. Together, the multiple explorations with AI enable Surrealism Me to offer new possibilities and a wider range of choices beyond the physical capabilities of the user's body.

\subsection{The World in Mixed Reality}

To enable the user to place their virtual body in MR, we propose a crowd-sourcing strategy to continuously map the world to acquire the technical infrastructure of MR. In Surrealism Me, we employ multiple techniques to empower the user with the capability to register and map the real world, allowing for the integration of virtual content, such as virtual body, seamlessly into the world. This approach enables the user to select and acquire Areas of Interest (AOI) and position the virtual body within these AOIs in the MR environment.

\subsubsection{MR Solution}
  
We employ Immersal SDK as the MR solution and integrate the Immersal Cloud Service to construct 3D point-clouds of selected AOI (see Fig.\ref{fig:MRtech-a}) using user-uploaded images. These point-clouds work with Visual Positioning System (VPS) to provide high-accuracy localization and identification of the AOI (in Fig.\ref{fig:MRtech-b}). The combination of local SLAM and VPS creates the illusion that virtual content is attached to the real world (see Fig.\ref{fig:MRtech-c}). 
  

\begin{figure}[b]
  \centering
  \begin{subfigure}{0.48\textwidth}
      \includegraphics[width=\textwidth]{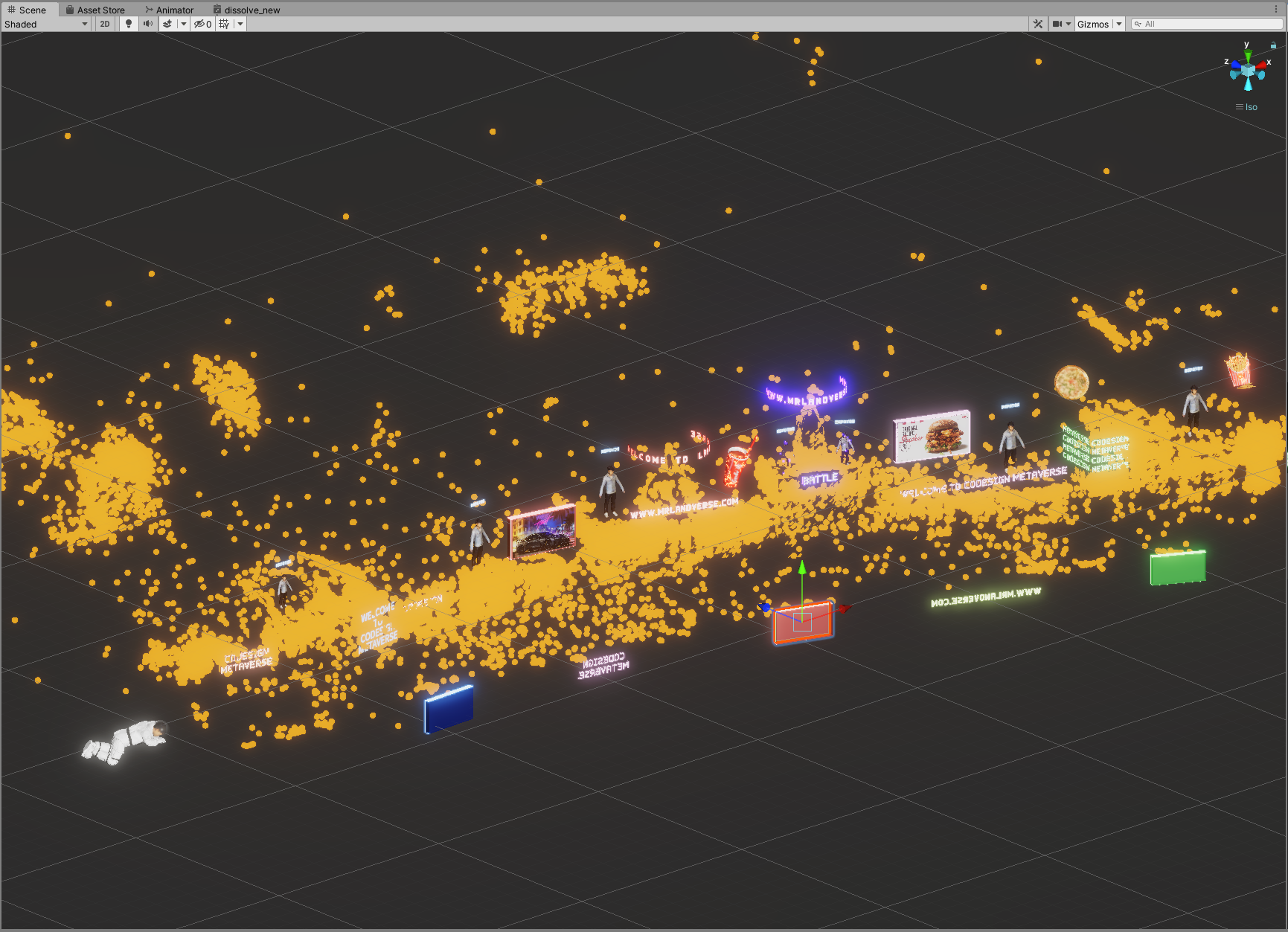}
      \caption{Interface to edit virtual content.}
      \label{fig:place-left}
  \end{subfigure}
  \hfill
  \begin{subfigure}{0.495\textwidth}
      \includegraphics[width=\textwidth]{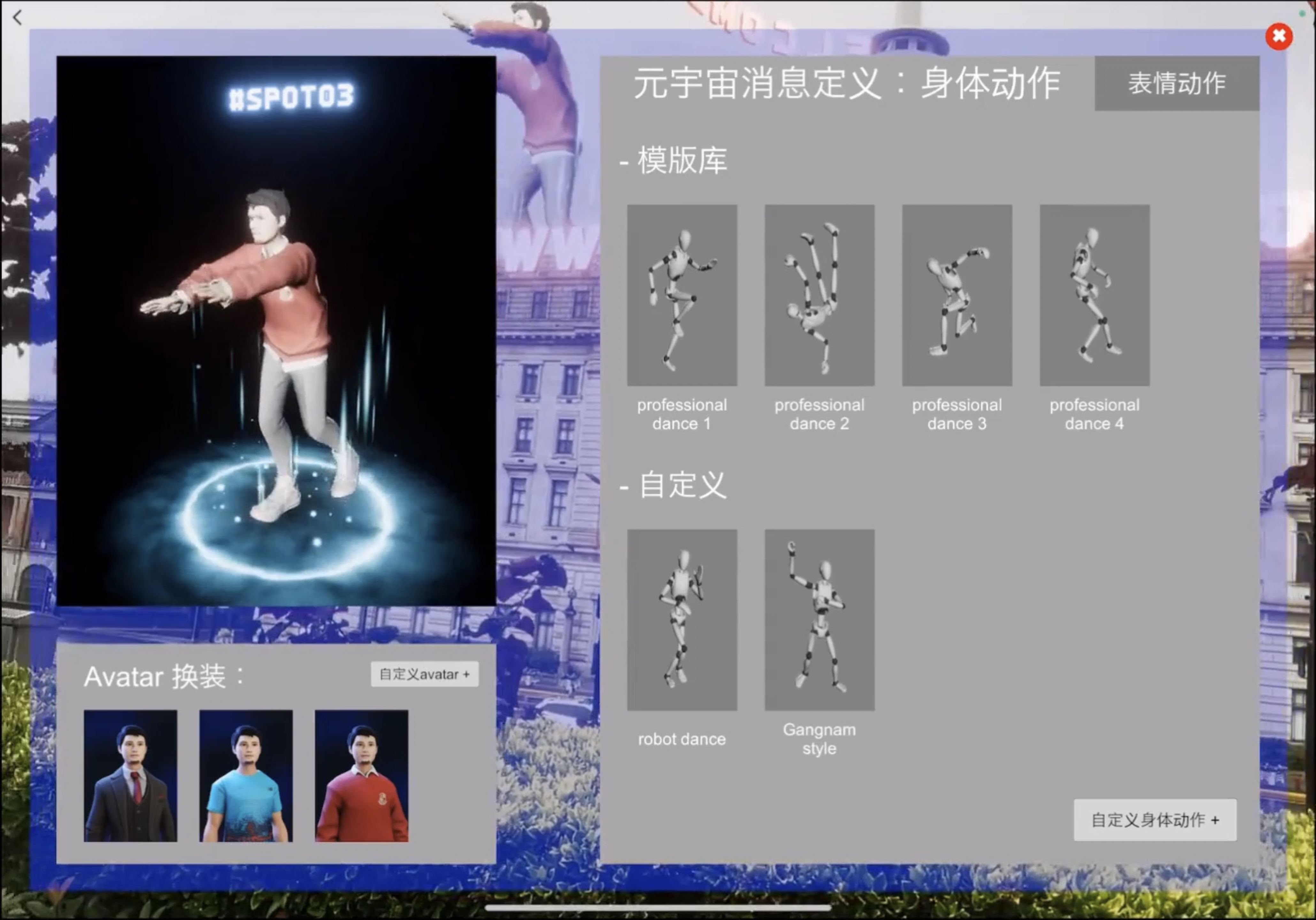}
      \caption{Interface to select motion from the database.}
      \label{fig:place-right}
  \end{subfigure}
  \caption{Placement of the virtual body (to MR).}
  \label{fig:place}
  \end{figure}
  
\begin{figure}
  \centering
  \includegraphics[width=\textwidth]{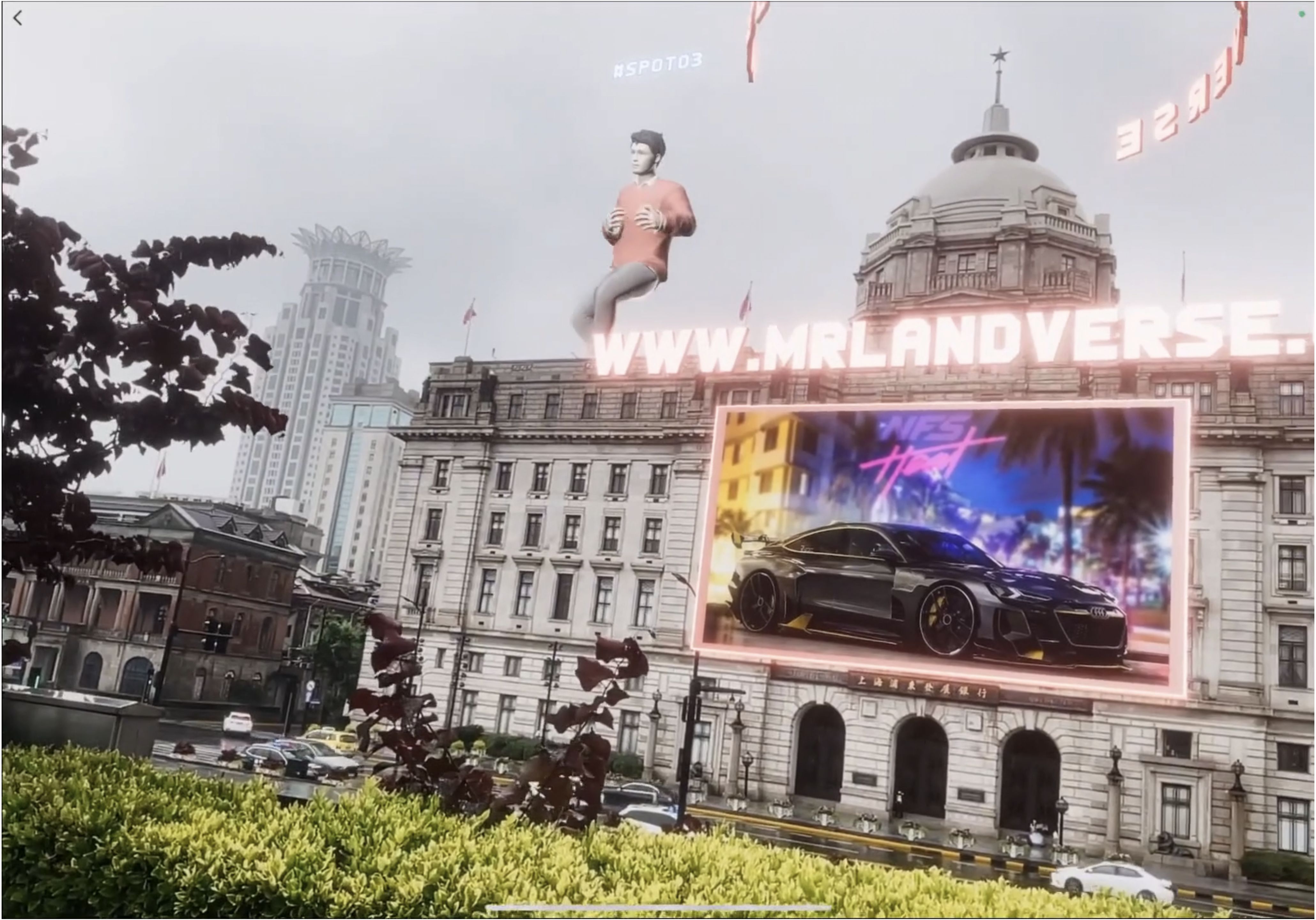}
  \caption{Virtual body in the MR environment. Screenshot of Surrealism Me.}
  \label{fig:place-demo}
  \end{figure}

  \subsubsection{Create New Site}
Surrealism Me offers a “Create New Site” function (in Fig.\ref{fig:map-a}) to crowd-source the MR infrastructure. The user can take a sequence of photos to map an AOI into MR (Fig.\ref{fig:map-b}) with two effective methods (see Fig.\ref{fig:map-c}), to break the AOI into multiple spots and encircle each to capture edge-overlapping photos, and to take two rounds of photos in left and right 45-degree angles. Surrealism Me automatically processes the photos into point-clouds to anchor virtual content. This approach allows the user to collectively gather and harness preferred locations and even the entire world to enclose their virtual body.


\subsubsection{Place the Virtual Body}

Surrealism Me integrates distinct interactions to enable the user to place the virtual body within registered AOIs. Fig.\ref{fig:place-left} showcases how to freely add and manipulate the position and scale of the virtual body. Fig.\ref{fig:place-right} demonstrates the interface to navigate through the motion database for an animated virtual body and place to the AOI. Consequently, this creates the perceptual illusion of the (virtual) body's presence within the MR environment (in Fig.\ref{fig:place-demo}). Our demonstration scene includes additional "decorative" elements intended for a visually captivating presentation and other research purposes, which are beyond the investigation of this paper, thus not discussed. 



\subsection{Sense of Embodiment (SOE) Preservation}

\begin{figure}[h]
  \includegraphics[width=\textwidth]{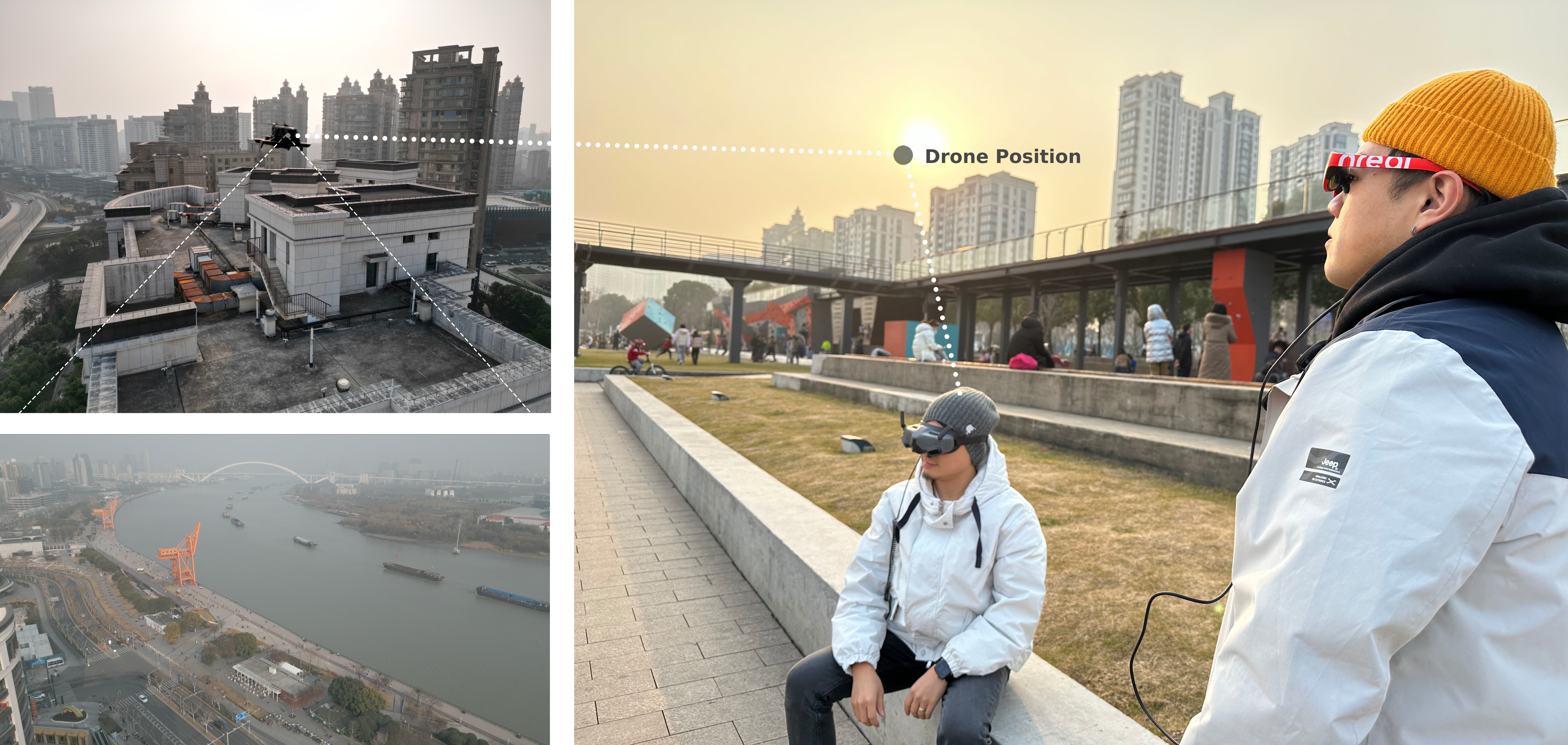}
  \caption{Embodied visionary experiences of the virtual body.}
  \label{fig:vision}
  \end{figure}

The feeling of having a (virtual) body goes beyond just manipulating the bodily movement. Perceiving other sensations, such as vision, and the feeling of being inside the body play a significant role \cite{6797786}. Surrealism Me enables one to manipulate and see their virtual body in the MR environment; the experience of having two bodies are present, yet the feeling of owning the virtual body is insufficient. To address this, we tend to synchronize the real time vision of the virtual body to the user. By automatically receiving the coordinates of the virtual body, the UAV autonomously reaches real-world destinations and acts as the eye of the virtual body. Through a first-person view (FPV), the goggle synchronizes with the UAV's camera (see Fig.\ref{fig:vision}). It lets the user experience the real-time vision of the virtual body, creating a heightened sense of being inside the virtual body. 

\begin{figure}[h]
  \includegraphics[width=\textwidth]{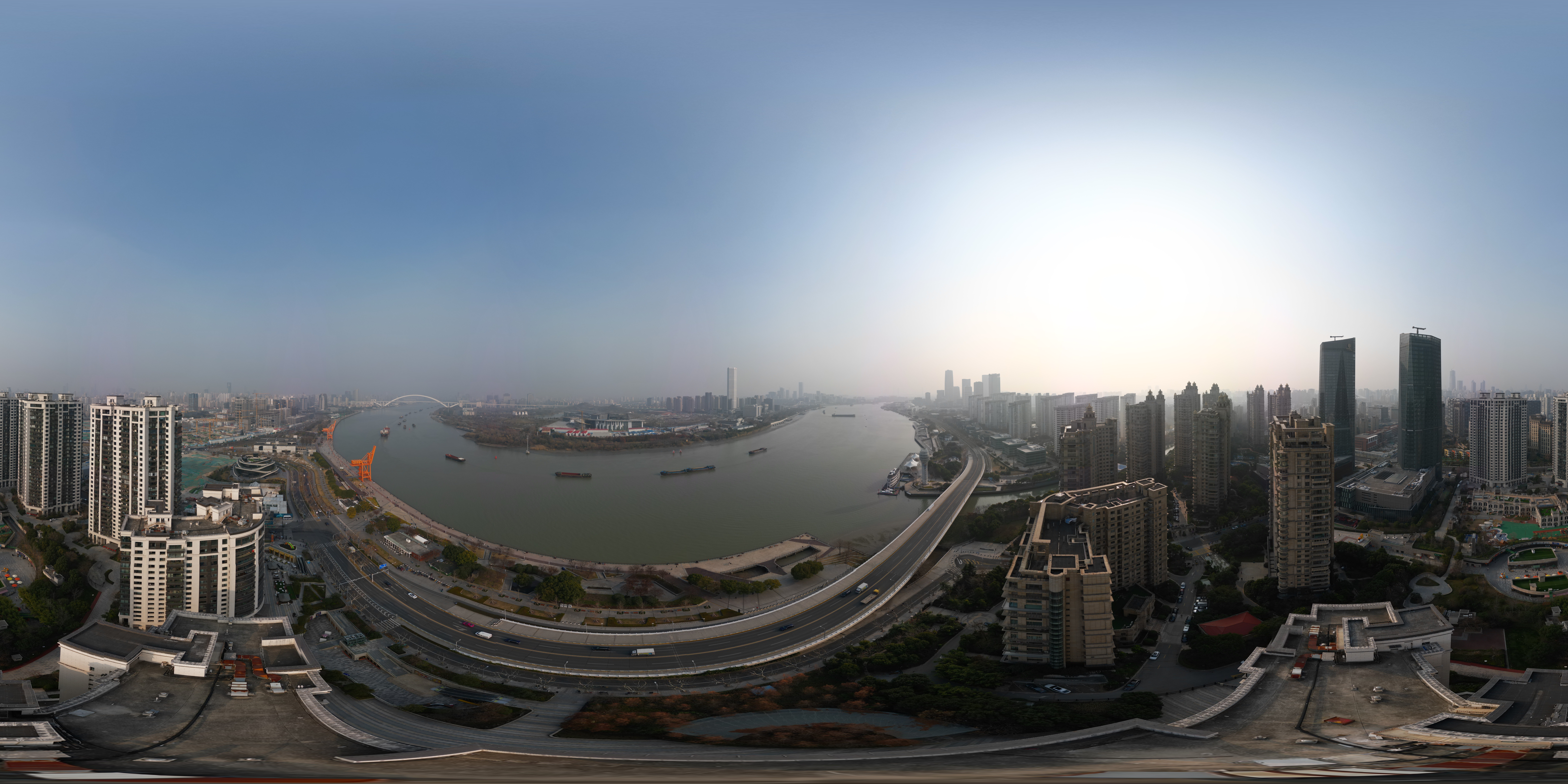}
  \caption{FPV view of Real-time 360-degree image.}
  \label{fig:drone}
  \end{figure}

Seeing through the virtual body further contributes to the embodied experiences in MR. The goggle allows the user to (optionally) control the perspective of the "flying eye" in real time, enhancing the embodied experiences. It also superficially touches on the idea of "playing against apparatus" through the interaction. Combined with the MR experiences, these elements preserve the sense of embodiment and enhance the ownership of the virtual body. The captured 360-degree view is accessible via headset and an interactive website (in Fig.\ref{fig:drone}) in both real time and afterwards. The camera projects the world in a different manner in these experiences and serves our intention of alienating the user's view from their ordinary perspective. 

\section{Related Work and Discussion}

As discussed in Section \ref{anotherbody}, various previous works investigate Sense of Embodiment (SOE) in VR and AR \cite{genay2021virtual, 9495125, 7504682}. In research related to SOE and MR, some works primarily investigate gaming technology and others focus on telepresence or the embodied experiences across realities \cite{9319120, 10.1145/3487983.3488304, 8466636}. All these works emphasize simulation of the world in MR, which indeed contain the misperception that confuses the user about the relationship between the media and the world. Through Surrealism Me, we investigate how humans can realize the obscuration of the world in MR and better understand the media.


\subsection{A Play of Humans Against Apparatus}

The growing motion database and its role in Surrealism Me imply a play of humans against apparatus, similar to the interaction between player verse game. The program of AI in Surrealism Me evolves dynamically with the user-generated motion data as new input and training data. The user continuously performs new movements that are not yet included in the program (i.e., unpredictable movement to the apparatus) and expands the motion database. Surrealism me constantly use the new database to train the motion generation algorithm (i.e., the AI neural network) and update the AI program. In other words, unpredictable intentions of humans impose the program of AI to evolve. On this note, the AI apparatus keeps updates itself towards ideal conditioned by the user. 

This program of AI embraces the unpredictable intentions of humans from the very beginning. An ideal AI for motion-capturing and generation could master all human movements. Only after AI has been ceaselessly transcended by human movements it had not contained, can it update itself to become ideal. Accordingly, the user has to continuously perform movements that are not yet included in the program, which demonstrates the user's unpredictable intentions in playing against apparatus and not merely a function of it but conditioned on it.

\subsection{An Ambivalence to Rethink the Reality of MR}

In Surrealism Me, seeing through the perspective of the virtual body creates an ambivalence: it offers strong embodied experiences but causes an alienation from the experiences of everyday life. This ambivalence inspires the user to rethink the reality of MR. When wearing the FPV goggle, the user expects the visionary feeling of the virtual body as an enhancement to embodied experiences. The FPV goggle livestreams the UAV's camera to the user, meanwhile the user can turn their head at will to control the UAV's perspective; a real-time dynamic panoramic view and enhanced embodied experiences are presented—if not taking into account the unusual height of the UAV's position. In other words, the user has immersive experiences of viewing the world directly in four dimensions, as to being inside the virtual body.

On the other hand, the UAV is often positioned at a high altitude when mimicking the perspective of the giant virtual body, such a viewing is not usually accessible to the user. It creates an alienated feeling especially when the user recognizes that these images flying between or above buildings are not what they see in everyday life, even though they look realistic. This unreal feeling contrasts with the enhanced, immersive embodied experiences of a four-dimensional world, which further heightens a sense of alienation. When the user contemplates the cause of this alienation, it is easy to aware of the presence of UAV's camera. Thus, they can realize that the seemingly real world in MR is not real, but a world projected by the apparatus.

\section{Conclusion}

Taking the artwork Surrealism Me as an example, this paper responds to Flusserian freedom in media studies by enabling the user to have a virtual body and enhanced embodied experiences. It first achieves a play against apparatus (i.e. AI), in which the user's unpredictable intentions are important, and then creates an alienated feeling of the unreal reality in MR. The user and the AI together exhaust the potentials of the program, thus leading to its evolution. The ambivalence between the embodied experiences and the high-altitude perspective of the virtual body reinforces the user's sense of alienation, stimulating the awareness that the world projected by MR is not real. Both reveal the possibilities of going beyond the domination of apparatus and approaching Flusserian Freedom in the contemporary technical arrangements.

\bibliographystyle{ACM-Reference-Format}
\bibliography{sample-authordraft}

@String{Computing = "Computing" }

@String{Computer = "{IEEE} Computer" }

@ARTICLE{9495125,
  author={Genay, Adélaïde and Lécuyer, Anatole and Hachet, Martin},
  journal={IEEE Transactions on Visualization and Computer Graphics}, 
  title={Being an Avatar “for Real”: A Survey on Virtual Embodiment in Augmented Reality}, 
  year={2022},
  volume={28},
  number={12},
  pages={5071-5090},
  doi={10.1109/TVCG.2021.3099290}}

@ARTICLE{6797786,
  author={Kilteni, Konstantina and Groten, Raphaela and Slater, Mel},
  journal={Presence}, 
  title={The Sense of Embodiment in Virtual Reality}, 
  year={2012},
  volume={21},
  number={4},
  pages={373-387},
  doi={10.1162/PRES_a_00124}}

@INPROCEEDINGS{7504682,
  author={Argelaguet, Ferran and Hoyet, Ludovic and Trico, Michael and Lecuyer, Anatole},
  booktitle={2016 IEEE Virtual Reality (VR)}, 
  title={The role of interaction in virtual embodiment: Effects of the virtual hand representation}, 
  year={2016},
  volume={},
  number={},
  pages={3-10},
  doi={10.1109/VR.2016.7504682}}

@article{genay2021virtual,
  title={Virtual, real or mixed: How surrounding objects influence the sense of embodiment in optical see-through experiences?},
  author={Genay, Ad{\'e}la{\"\i}de and L{\'e}cuyer, Anatole and Hachet, Martin},
  journal={Frontiers in Virtual Reality},
  volume={2},
  pages={679902},
  year={2021},
  publisher={Frontiers Media SA}
}

@book{flusser2000,
  title={Towards a philosophy of photography},
  author={Flusser, Vil{\'e}m},
  year={2013},
  publisher={Reaktion Books}
}

@article{flusser2013,
  title={Our Images},
  author={Flusser, Vil{\'e}m},
  journal={Post-history, Minneapolis: Univocal Publishing},
  year={2013}
}

@article{flusser1997,
  title={Nachgeschichte: eine korrigierte Geschichtsschreibung},
  author={Flusser, Vil{\'e}m},
  year={1993}
}

@book{flusser2011,
  title={Into the universe of technical images},
  author={Flusser, Vil{\'e}m},
  volume={32},
  year={2011},
  publisher={U of Minnesota press}
}

@book{flusser2002,
  title={Medienkultur: Die Bildwelten der Informationsgesellschaften},
  author={Flusser, Vil{\'e}m},
  year={2002},
  publisher={Fischer Taschenbuch Verag}
}

@article{ieven2003,
  title={How to orientate oneself in the world: A general outline of Flusser’s theory of media},
  author={Ieven, Bram},
  journal={Image \& Narrative},
  volume={6},
  year={2003}
}

@book{poltronieri2014communicology,
  title={Communicology, apparatus, and post-history: Vil{\'e}m flusser’s concepts applied to video games and gamification},
  author={Poltronieri, Fabrizio},
  year={2014},
  publisher={meson press}
}

@book{popiel2012vilem,
  title={Vil{\'e}m Flusser's media philosophy: Tracing the digital in nature through art},
  author={Popiel, Anne Marie},
  year={2012},
  publisher={Washington University in St. Louis}
}

@online{unityAR,
	title = {{AR} Foundation {\textbar} {AR} Foundation {\textbar} 5.1.1},
	url = {https://docs.unity3d.com/},
	author = {Unity Software Inc.},
	urldate = {2024-01-09},
  year={2022},
}

@online{mixamo,
	title = {Mixamo},
	url = {https://www.mixamo.com/#/},
  author = {Mixamo Inc.},
	urldate = {2024-01-09},
  year={2022},

}

@ARTICLE{9042236,
  author={Sun, Guofei and Wong, Yongkang and Cheng, Zhiyong and Kankanhalli, Mohan S. and Geng, Weidong and Li, Xiangdong},
  journal={IEEE Transactions on Multimedia}, 
  title={DeepDance: Music-to-Dance Motion Choreography With Adversarial Learning}, 
  year={2021},
  volume={23},
  number={},
  pages={497-509},
  doi={10.1109/TMM.2020.2981989}}

@INPROCEEDINGS{9880214,
  author={Guo, Chuan and Zou, Shihao and Zuo, Xinxin and Wang, Sen and Ji, Wei and Li, Xingyu and Cheng, Li},
  booktitle={2022 IEEE/CVF Conference on Computer Vision and Pattern Recognition (CVPR)}, 
  title={Generating Diverse and Natural 3D Human Motions from Text}, 
  year={2022},
  volume={},
  number={},
  pages={5142-5151},
  doi={10.1109/CVPR52688.2022.00509}}

@INPROCEEDINGS{9319120,
  author={Chung, Yu-Yen and Guo, Hung-Jui and Kumar, Hiranya Garbha and Prabhakaran, Balakrishnan},
  booktitle={2020 IEEE International Conference on Artificial Intelligence and Virtual Reality (AIVR)}, 
  title={High-quality First-person Rendering Mixed Reality Gaming System for In Home Setting}, 
  year={2020},
  volume={},
  number={},
  pages={339-344},
  doi={10.1109/AIVR50618.2020.00070}}

@inproceedings{10.1145/3487983.3488304,
author = {Abbey, Alexandre and Porssut, Thibault and Herbelin, Bruno and Boulic, Ronan},
title = {Assessing the Impact of Mixed Reality Immersion on Presence and Embodiment},
year = {2021},
isbn = {9781450391313},
publisher = {Association for Computing Machinery},
address = {New York, NY, USA},
url = {https://doi.org/10.1145/3487983.3488304},
doi = {10.1145/3487983.3488304},
booktitle = {Proceedings of the 14th ACM SIGGRAPH Conference on Motion, Interaction and Games},
articleno = {2},
numpages = {10},
location = {Virtual Event, Switzerland},
series = {MIG '21}
}

@ARTICLE{8466636,
  author={Piumsomboon, Thammathip and Lee, Gun A. and Ens, Barrett and Thomas, Bruce H. and Billinghurst, Mark},
  journal={IEEE Transactions on Visualization and Computer Graphics}, 
  title={Superman vs Giant: A Study on Spatial Perception for a Multi-Scale Mixed Reality Flying Telepresence Interface}, 
  year={2018},
  volume={24},
  number={11},
  pages={2974-2982},
  doi={10.1109/TVCG.2018.2868594}}

\appendix \label{avatar}


\section{The user self-created avatar}

\begin{figure}[b]
  \centering
  \begin{subfigure}{\textwidth}
      \includegraphics[width=\textwidth]{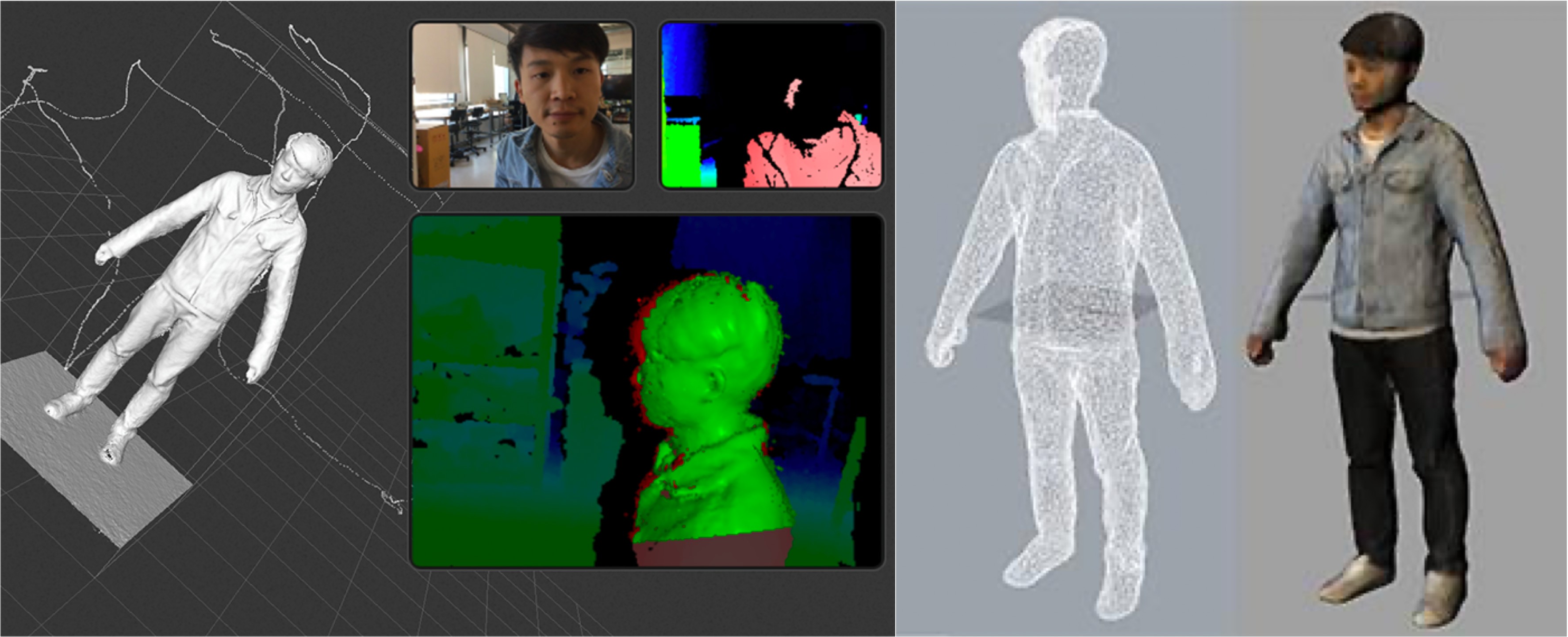}
       \caption{}
      \label{fig:scan}
  \end{subfigure}
  \hfill
  \begin{subfigure}{\textwidth}
      \includegraphics[width=\textwidth]{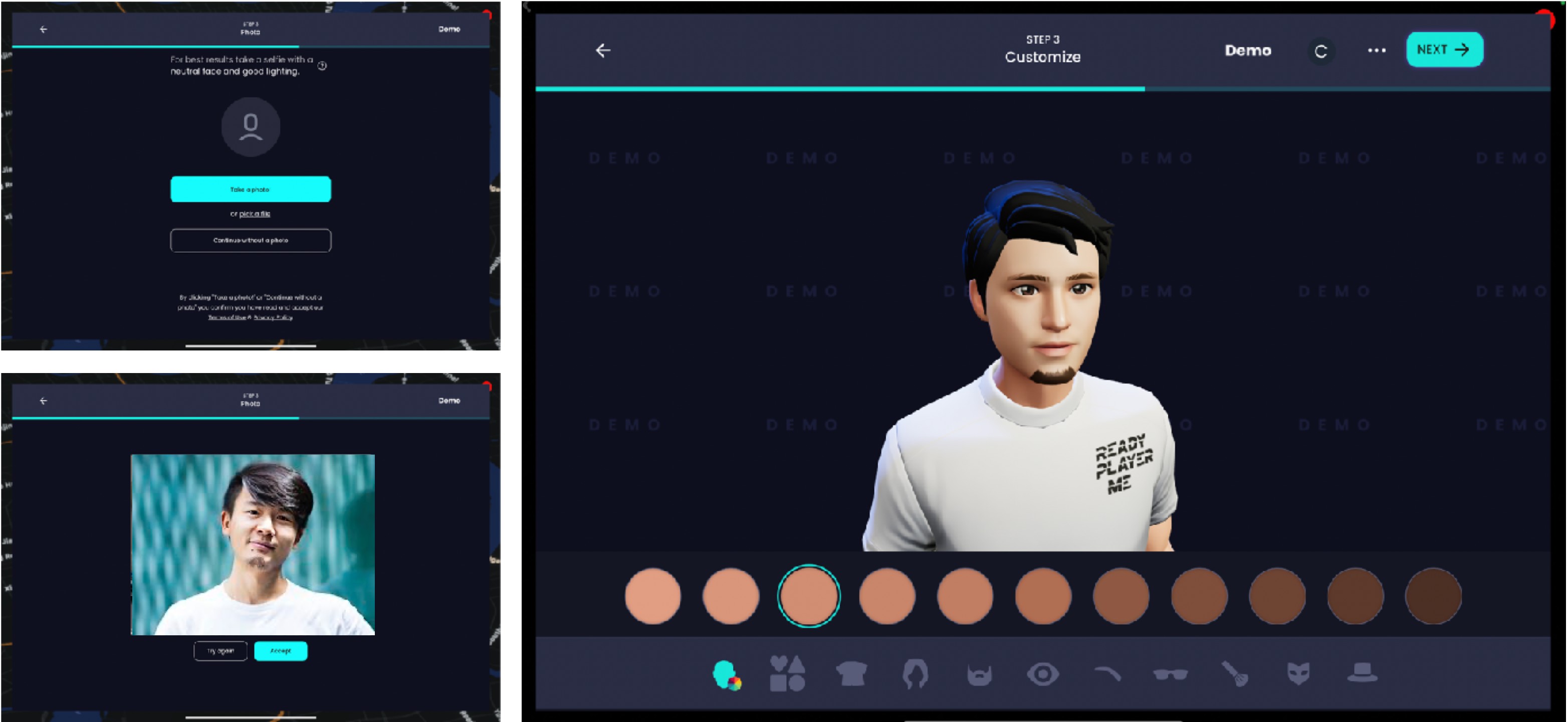}
      \caption{}
      \label{fig:palyer}
  \end{subfigure}
  \caption{Interface to create avatar for the virtual body.}
  \label{fig:avatarfig}
  \end{figure}

Surrealism Me provides two different method to the user to easily obtain their avatar as the virtual body. Fig.\ref{fig:scan} demonstrates the workflow to scan the user as their avatar. Fig.\ref{fig:palyer} shows the integrated function in Surrealism Me (base on Ready Player SDK) and its interface allow the user to customize their avatar.








\end{document}